\documentclass[twocolumn,english,aps,pra,superscriptaddress]{revtex4-2}
\usepackage[T1]{fontenc}
\usepackage[utf8]{inputenc}
\usepackage{babel}
\usepackage{amsmath}
\usepackage{amsthm}
\usepackage{graphicx}
\usepackage{subscript}
\usepackage[unicode=true,pdfusetitle,
 bookmarks=true,bookmarksnumbered=false,bookmarksopen=false,
 breaklinks=false,pdfborder={0 0 1},backref=false,colorlinks=false]
 {hyperref}
\hypersetup{
 colorlinks=true}

\makeatletter
\numberwithin{equation}{section}
\numberwithin{figure}{section}

\usepackage{units}
\usepackage{babel}
\usepackage{float}
\usepackage{graphicx}
\usepackage{babel}
\usepackage{enumitem}
\renewcommand{\theenumi}{\alph{enumi}}
\usepackage{tikz}
\usepackage{vmargin}
\setmargins{1.5cm}       
{0.5cm}                        
{17.5cm}                      
{25.42cm}                    
{14pt}                           
{1cm}                           
{0pt}                             
{0cm}                           
\usepackage{amsmath}
\usepackage{amssymb}
\usepackage{cancel}
\usepackage[nodayofweek,yyyymmdd]{datetime}
\usepackage{mdframed}
\definecolor{green}{RGB}{0,128,0}
\renewcommand{\fnum@figure}[1]{\textup{\textbf{FIG. \thefigure}}\textup{:} \upshape}  

\renewcommand\thefigure{\arabic{figure}}

\fontsize{9pt}{10pt}\selectfont 
\usepackage[utf8]{inputenc}

\makeatother

\begin{document}
\fontsize{9pt}{10pt}\selectfont 
\title{Enhancing the ODMR Signal of Organic Molecular Qubits}
\author{\normalsize Yong Rui Poh}
\email{ypoh@ucsd.edu}

\affiliation{Department of Chemistry and Biochemistry, University of California
San Diego, La Jolla, California 92093, USA}
\author{\normalsize Joel Yuen-Zhou}
\email{joelyuen@ucsd.edu}

\affiliation{Department of Chemistry and Biochemistry, University of California
San Diego, La Jolla, California 92093, USA}
\date{November 20, 2024}
\begin{abstract}
In quantum information science and sensing, electron spins are often
purified into a specific polarisation through an optical-spin interface,
a process known as optically-detected magnetic resonance (ODMR). Diamond-NV
centres and transition metals are both excellent platforms for these
so-called colour centres, while metal-free molecular analogues are
also gaining popularity for their extended polarisation lifetimes,
milder environmental impacts, and reduced costs. In our earlier attempt
at designing such organic high-spin $\pi$-diradicals, we proposed
to spin-polarise by shelving triplet $M_{S}=\pm1$ populations as
singlets. This was recently verified by experiments albeit with low
ODMR contrasts of $<1\%$ at temperatures above 5 K. In this work,
we propose to improve the ODMR signal by moving singlet populations
back into the triplet $M_{S}=0$ sublevel, designing a true carbon-based
molecular analogue to the NV centre. Our proposal is based upon transition-orbital
and group-theoretical analyses of beyond-nearest-neighbour spin-orbit
couplings, which are further confirmed by ab initio calculations of
a realistic trityl-based radical dimer. Microkinetic analyses point
towards high ODMR contrasts of around $30\%$ under experimentally-feasible
conditions, a stark improvement from previous works. Finally, in our
quest towards ground-state optically-addressable molecular spin qubits,
we exemplify how our symmetry-based design avoids Zeeman-induced singlet-triplet
mixings, setting the scene for realising electron spin qubit gates.
\end{abstract}
\maketitle
\global\long\def\blue#1{\textcolor{blue}{#1}}%
\global\long\def\red#1{\textcolor{red}{#1}}%
\global\long\def\green#1{\textcolor{green}{#1}}%
\global\long\def\purple#1{\textcolor{purple}{#1}}%
\global\long\def\orange#1{\textcolor{orange}{#1}}%

\global\long\def\it#1{\textit{\textrm{#1}}}%
\global\long\def\un#1{\underline{\textrm{#1}}}%
\global\long\def\br#1{\left( #1 \right)}%
\global\long\def\sqbr#1{\left[ #1 \right]}%
\global\long\def\curbr#1{\left\{  #1 \right\}  }%
\global\long\def\braket#1{\langle#1 \rangle}%
\global\long\def\bra#1{\langle#1 \vert}%
\global\long\def\ket#1{\vert#1 \rangle}%
\global\long\def\abs#1{\left|#1\right|}%
\global\long\def\mb#1{\mathbf{#1}}%
\global\long\def\doublebraket#1{\langle\langle#1 \rangle\rangle}%

\section*{Introduction}

\begin{figure*}
\emph{\includegraphics[width=1\textwidth]{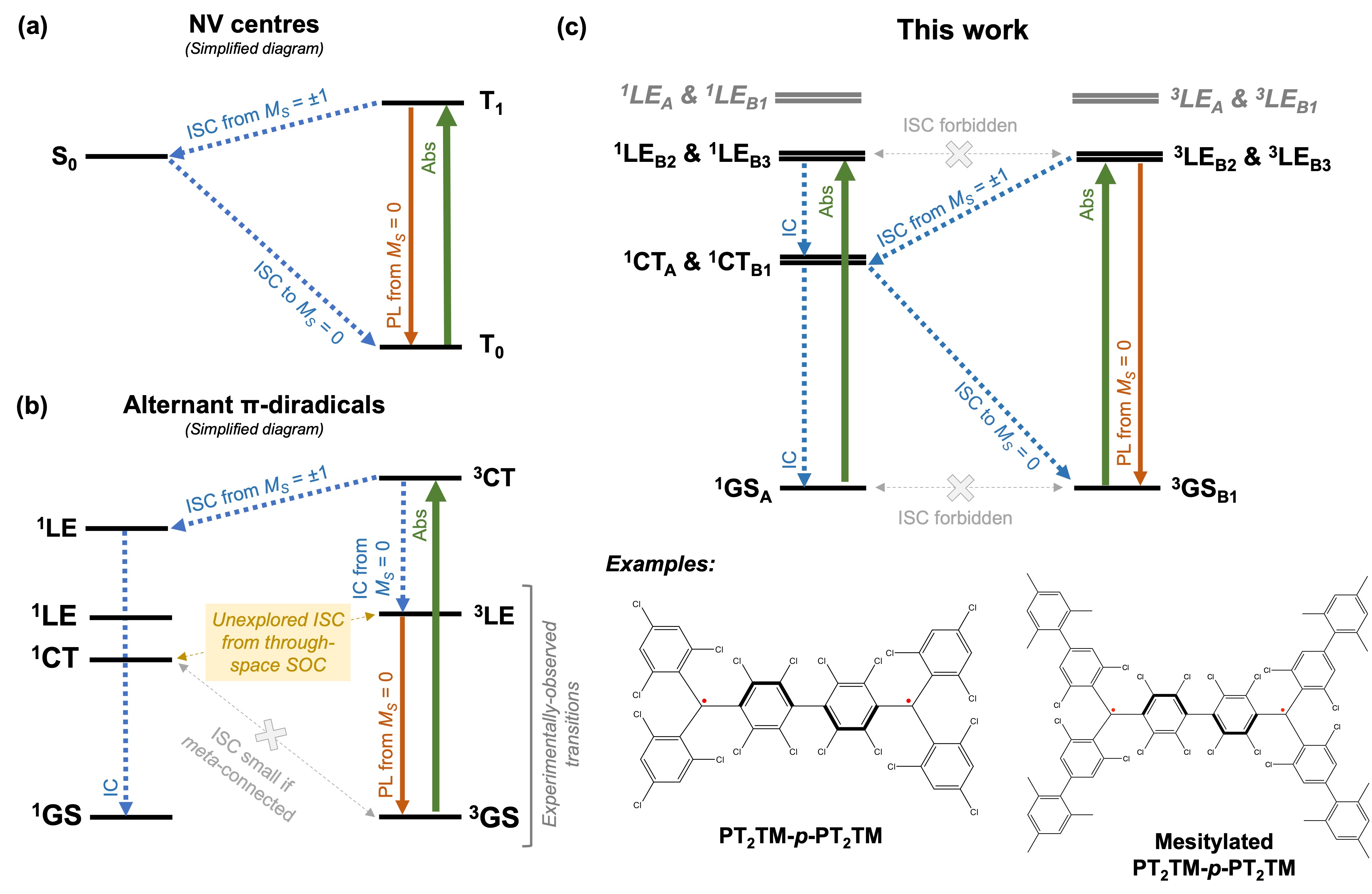}\caption{\label{fig:main}(a) Simplified ODMR mechanism typical of diamond-NV
defects. (b) Simplified ODMR mechanism of alternant $\pi$-diradicals
as explored by our previous work \citep{Poh2024}. Note that the CT-to-GS
ISC rates are small if the two benzylic radicals are tethered at their
\emph{meta} positions, which is avoided by our new design (see main
text). (c) ODMR mechanism of our new design, which more closely resembles
the spin polarisation pathway of NV centres. Also drawn are the PT\protect\textsubscript{2}TM-\emph{p}-PT\protect\textsubscript{2}TM
diradical and its mesitylated counterpart, both of which satisfy our
design requirements. (Abbreviations of electronic transitions: Abs
= absorption; PL = photoluminescence; IC = internal conversion; ISC
= intersystem crossing. Labels for electronic states: GS = ground
state; LE = local excitation; CT = charge transfer. These labels are
inherited from our previous analysis of the $\pi$-diradical electronic
structure \citep{Poh2024} with subscripts $A$, $B_{1}$, $B_{2}$,
and $B_{3}$ signifying the state irreps.)}
}
\end{figure*}

Polarised electron spins are promising quantum bit (qubit) candidates
\citep{Awschalom2018} with applications in quantum sensing \citep{Abobeih2019}
and quantum information science \citep{Pfaff2014}. Purifying these
spin magnetic dipoles into a particular polarisation requires irreversible
decays and these have been achieved in solid-state spin defects \citep{Taylor2008,Degen2017,Rose2018,Gottscholl2020,Chejanovsky2021,Mukherjee2023,Li2024,Dreau2011,Tetienne2012,Li2024-2}
by coupling the microwave-addressable electron spins with their orbital
degrees of freedom {[}Fig. \ref{fig:main}a{]}. Because the latter
lies high in the optical regime, this technique has become known as
optically-detected magnetic resonance (ODMR) -- ``detected'' because
the same optical-spin interface also enables readout of the spin polarisation.
Platforms hosting optically addressable spins are then referred to
as colour centres.

While early endeavours in this direction have focused on solid-state
defects such as nitrogen-vacancy (NV) centres in diamond \citep{Doherty2013},
their poor scalability and tunability have motivated the community
to shift towards molecular platforms \citep{GaitaArino2019,Atzori2019,Wasielewski2020,Yu2021,Laorenza2022,Scholes2023,Wu2023,Yang2016,Huang2023,Zhou2024a}.
Transition metal complexes with high-spin ground states (GSs) offer
the most natural starting point and this has been successful in Cr(IV),
V(III), Ni(II), and, most recently, Ir(IV) complexes \citep{Wojnar2020,Bayliss2020,Fataftah2020,Mirzoyan2021,Kazmierczak2021,Laorenza2021,Amdur2022,Bayliss2022,Goh2022,Kazmierczak2022,Kazmierczak2023,Mullin2023,Sutcliffe2024}.
However, in none of these systems were the spin lifetimes as long
as NV centres, essential for applications like qubit operations, and
this has been attributed to the larger spin-orbit couplings (SOCs)
introduced by metals. Other challenges include higher costs and poorer
sustainability. This has prompted a search for a molecular NV-centre
mimic that is metal-free, just like diamond itself, notwithstanding
the benefits of metal-based qubit systems such as higher biocompatibility
\citep{Ihara2019,Mishra2023} and simpler protocols for multiqubit
addressability \citep{Aguila2014,Moreno2018,Atzori2019,Moreno2021}.

As a first step along this path, organic molecules were designed to
demonstrate ODMR in high-spin excited states while acknowledging the
limited lifetimes of these electronic excitations \citep{Smyser2020,Dill2023,Gorgon2023,Palmer2024,Mena2024,Singh2024,Privitera2024,Yamauchi2024,Lin2024}.
Subsequently, our group theoretically proposed a class of organic
$\pi$-diradicals with optically-addressable high-spin ground states
\citep{Poh2024} that were also experimentally validated by Chowdhury
et al. \citep{Chowdhury2024} and Kopp et al. \citep{Kopp2024}. In
all of these systems, spin polarisation was likely attained by shelving
the triplet $M_{S}=\pm1$ population in the singlet manifold via spin-selective
intersystem crossings (ISCs) while utilising the presence of additional
singlet charge-recombined states absent from the triplet manifold
{[}Fig. \ref{fig:main}b{]}. (Interestingly, such imbalances have
recently been proposed as evidence of low-energy electronic spin isomers
\citep{Shimizu2024}.) Missing from the ODMR cycles are the ISCs between
singlets and triplets selective for the $M_{S}=0$ triplet sublevel,
which the NV centre exhibits. As mentioned in our previous work \citep{Poh2024},
this limits the maximum possible triplet $M_{S}=0$ population to
$25\%$ at steady state, constraining the ODMR resolution. Moreover,
this increases the risk of metastable singlet molecules returning
as $M_{S}=\pm1$ triplets, which would negate the accumulated spin
polarisation. Indeed, in the aforementioned experimental works, poor
ODMR contrasts of $<1\%$ were observed at temperatures above 5 K.
We note in passing that, apart from $\pi$-diradicals, alkaline earth
metal complexes \citep{Khvorost2024,Wojcik2024} and nitrenes \citep{Gately2023}
also display diradical properties and their functions as molecular
colour centres have not been fully explored.

Our earlier theoretical model \citep{Poh2024} had focused primarily
on nearest-neighbour-only (through-bond-only) SOCs, which predicted
ISCs to have $\Delta M_{S}=0$ ($\pm1$) selectivity for electron
spins quantised parallelly (perpendicularly) to the dimer linkage.
However, beyond-nearest-neighbour SOC effects have been shown to be
experimentally relevant in ODMR, as in the recent work by Kopp et
al. \citep{Kopp2024} {[}Fig. \ref{fig:main}b{]}, and such effects
can induce ISCs with an opposite spin selectivity. Motivated by this
observation, in this work, we design another class of organic $\pi$-diradicals,
this time capable of ISCs involving both triplet $M_{S}=\pm1$ and
$M_{S}=0$ sublevels of \emph{different} electronic states as facilitated
by beyond-nearest-neighbour and nearest-neighbour SOCs respectively
{[}Fig. \ref{fig:main}c{]}. This way, we create a true analogue of
the NV centre, which we predict to demonstrate enhanced ODMR signals
by optically pumping the near-degenerate triplet and singlet local
excitations. Our design principle stems from a careful group-theoretical
analysis of the transitioning orbitals, which coincidentally also
solves the problem of singlet-triplet mixing during electron paramagnetic
resonance (EPR) measurements of weakly-coupled diradicals (as most
$\pi$-diradicals are \citep{Shu2023,Kopp2024}). As a realistic prototypical
example of our design, the electronic structure of PT\textsubscript{2}TM-\emph{p}-PT\textsubscript{2}TM
{[}Fig. \ref{fig:main}c{]} was shown at both density functional theory
(DFT) and multi-configurational (MC) levels of theory to demonstrate
an ODMR mechanism similar to the NV centre. The potential for signal
enhancement through this ODMR pathway is supported by microkinetic
analyses, where we found high ODMR contrasts of around $30\%$ to
be experimentally possible (as compared to $<1\%$ in earlier theoretical
\citep{Poh2024} and experimental \citep{Chowdhury2024,Kopp2024}
studies). In this regime, the steady-state triplet population exceeds
$25\%$, which is the upper bound of prior works \citep{Poh2024,Chowdhury2024,Kopp2024}.
While not explored by the present work, we expect further mesitylation
{[}Fig. \ref{fig:main}c{]} to improve the luminescence quantum yield
via geometrical relaxation \citep{Murto2023,Ghosh2024}, as exemplified
by Chowdhury et al. \citep{Chowdhury2024} to produce near-unity photoluminescence
yields. This paves the way towards better designs of optically-addressable
organic spins. Note that the alternacy symmetry of PT\textsubscript{2}TM-\emph{p}-PT\textsubscript{2}TM
is not important to our design, which should be separated from our
earlier work \citep{Poh2024}.

\section*{Results and Discussions}

\subsection*{Investigating the role of beyond-nearest-neighbour SOCs}

\begin{figure*}
\emph{\includegraphics[width=1\textwidth]{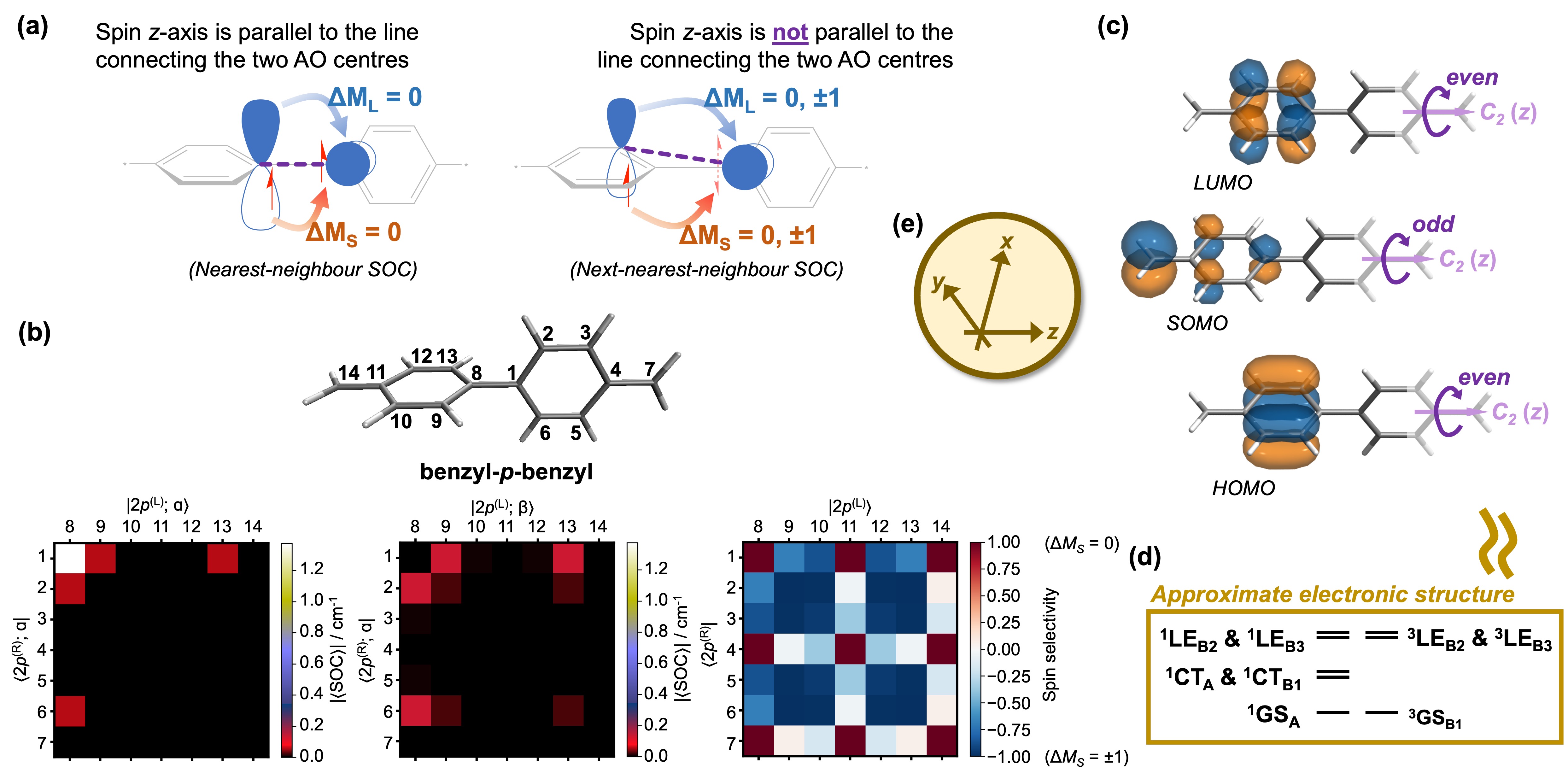}\caption{\label{fig:soc-benzyl}(a) Breakdown of the spin selection rule for
ISCs involving next-nearest-neighbour SOCs. (b) Amplitudes of the
SOC matrix elements evaluated between two carbon $2p$ AOs centred
on opposite $\pi$-systems of the benzyl-\emph{p}-benzyl diradical.
Each $2p$ AO is assumed to be occupied by one full electron with
either the same or opposite spin as its counterpart on the other benzylic
fragment. Taking the difference between the matrix elements-squared
of the two spin alignments and weighing the result by their sums give
the spin selectivity. (c) Sketches of the frontier Hückel MOs for
the benzylic monoradical. Also illustrated are their parities under
$C_{2}\protect\br z$ rotation. (d) Approximate electronic structure
of benzyl-\emph{p}-benzyl obtained from the Hückel theory analysis
in (c). The excitation labels are inherited from Ref. \citep{Poh2024}
and the subscripts indicate the state irreps. In this work, all molecular
geometries are presented in the coordinate frame depicted in (e).}
}
\end{figure*}

As we have previously shown \citep{Poh2024}, when two benzylic radicals
are covalently connected with a significant torsion (as is the way
most $\pi$-diradicals are experimentally constructed \citep{Hattori2019,Kimura2021,Wonink2021,Feng2021,Murto2022,Huang2022,Abdurahman2023,Schafter2023,Matsuoka2023,Liu2023,Prajapati2023,Abdurahman2023-2,Zhou2024,Chang2024,Mizuno2024,Wang2024,Yu2024,Shu2023}),
its SOC matrix elements are dominated by nearest-neighbour interactions
between the two $2p$ atomic orbitals (AOs) of opposite $\pi$-rings
(reason: El-Sayed rule \footnote{The El-Sayed rule states that the SOC matrix elements are the largest
when the transitioning orbitals adopt different orientations \citep{El-Sayed1963}.}) that are closest to the dimer linkage (reason: SOC is a local effect).
Henceforth, we refer solely to $\pi$ AOs. Furthermore, if the spin
quantisation axis lies parallel (perpendicular) to the line connecting
the two $2p$ AO centres, then the ISC spin selectivity arising from
the SOC matrix elements is strictly $\Delta M_{S}=0$ ($\pm1$) due
to symmetry \citep{Poh2024}. Note that the last condition is satisfied
for interactions at the dimer bond because the spin axes (wihout applying
any magnetic fields) coincide with the dimer's high symmetry axes.
However, ISCs in $\pi$-diradicals occur not between pairs of $2p$
AOs but between pairs of \emph{delocalised} molecular orbitals (MOs)
that are each a linear combination of a few $2p$ AOs. When one of
the transitioning MOs contains a node at the dimer linkage, the relevant
SOC matrix elements necessarily involve next-nearest-neighbour couplings
not considered by the above analysis. Such effects are generally small
following the local nature of SOCs, yet they can be appreciable if
most of the transitioning MO's amplitude is piled up at these next-nearest-neighbour
sites (as opposed to nearest-neighbour couplings weighted by small
MO amplitudes). Importantly, next-nearest-neighbour interactions involve
two $2p$ AO centres that are not colinear with the spin quantisation
axis, causing the ISC selectivity to break down {[}Fig. \ref{fig:soc-benzyl}a{]}.
Indeed, for two benzylic radicals covalently tethered at the \emph{para}
positions with $90^{\text{o}}$ torsion (hereafter called benzyl-\emph{p}-benzyl),
the next-nearest-neighbour SOC evaluated over singly-occupied $2p$
AOs biases the $\Delta M_{S}=\pm1$ transitions with $68.1\%$ selectivity
(noting that the spin quantisation ($z$-)axis lies parallel to the
dimer linkage due to the molecular $D_{2d}$ point group symmetry)
and this SOC has a significant amplitude of $0.13\mathrm{\,cm^{-1}}$
{[}Fig. \ref{fig:soc-benzyl}b{]}. For comparsion, the nearest-neighbour
analogue has a \emph{perfect} spin selectivity of $\Delta M_{S}=0$
with a larger amplitude of $1.37\mathrm{\,cm^{-1}}$. Note that these
SOC matrix elements represent upper bounds to the true molecular values
because we have assumed one full electron in each AO.

\subsection*{Perfecting the spin selectivity using the molecular point group symmetry}

That ISC in $\pi$-diradicals can actually occur with both spin selectivities
is contrary to most theoretical analyses \citep{Hong2001,Barford2010,Yu2012}
(including ours \citep{Poh2024}) and opens up the possibility of
transferring the full electronic structure of NV centres into molecules.
Our next step is to perfect the next-nearest-neighbour spin selectivity
via symmetry considerations. We note that by connecting two benzylic
radicals at their \emph{para} positions, the resulting diradical has
at least a $D_{2}$ point group symmetry with irreducible representations
(irreps) organised by their characters under $C_{2}\br y$ and $C_{2}\br z$
rotations. The irrep of any electronic state is in turn the direct
product of its orbital and spin irreps. Focusing on the orbital part,
if we continue to align the $z$-axis along the dimer linkage {[}Fig.
\ref{fig:soc-benzyl}b,e{]}, then a $C_{2}\br y$ rotation becomes
equivalent to swapping the two monomers. Therefore, for two weakly-interacting
$\pi$-systems of significant torsion, the low-lying excitations are
expected to appear in near-degenerate pairs of $\curbr{A,B_{1}}$
and $\curbr{B_{2},B_{3}}$, each comprising a state of opposite parity
under $C_{2}\br y$, i.e. equal probability of exciting either monomer.
These two pairs can be distinguished by their transformations under
$C_{2}\br z$ and, for that, it is easier to think in the basis of
\emph{monomeric} $\pi$ MOs. We first note that $2p$ AOs have $\pi$
symmetry, that is, their phases change under $C_{2}$ rotation about
the AO centre. Further, for the two $2p$ AOs at the dimer linkage,
that $\pi$-symmetry-defining $C_{2}$ rotation coincides with the
diradical's $C_{2}\br z$ symmetry element. Hence, a monomeric MO
with a non-zero amplitude at the dimer linkage \emph{must} be antisymmetric
under $C_{2}\br z$ rotation. By contrast, for a monomeric MO to be
symmetric under $C_{2}\br z$ rotation, it \emph{must} possess a node
at the dimer linkage. For instance, the Hückel method predicts the
highest occupied, singly occupied and lowest unoccupied MOs (abbreviated
as HOMO, SOMO, and LUMO respectively) of the benzylic radical to have
$+1$, $-1$, and $+1$ characters under $C_{2}\br z$ rotation {[}Fig.
\ref{fig:soc-benzyl}c{]}. Finally, we recall that the symmetry of
the orbital wavefunction is the direct product of the symmetries of
all singly occupied orbitals. Therefore, the ground states of benzyl-\emph{p}-benzyl,
having two unpaired electrons in the SOMOs, transform as either $A$
or $B_{1}$. Similarly, the lowest charge transfer (CT) singlets with
all doubly-occupied orbitals (also known as charge-recombined or zwitterionic
states of the two radicals) also transform as either $A$ or $B_{1}$.
As for the lowest local excitations (LEs), they are a linear combination
of HOMO-to-SOMO and SOMO-to-LUMO excitations (an outcome of alternancy
symmetry; see Ref. \citep{Poh2024}) and hence transform as either
$B_{2}$ or $B_{3}$. Including the characters under $C_{2}\br y$
rotation yields the approximate energy level diagram illustrated in
Fig. \ref{fig:soc-benzyl}d (for a review of the electronic structure
of benzylic diradicals, see our earlier work \citep{Poh2024}).

Returning to our goal of attaining perfect ISC spin selectivities,
we note that the singlet, triplet $M_{S}=0$, and triplet $M_{S}=\pm1$
spin sublevels transform as $A$, $B_{1}$, and $\curbr{B_{2},B_{3}}$
respectively \citep{DeGroot1967,Szumska2019}. For an ISC to be symmetry-allowed,
the SOC matrix elements must contain the totally symmetric irrep.
Hence, because the SOC Hamiltonian is totally symmetric, ISC processes
in benzyl-\emph{p}-benzyl are spin-selective for $\Delta M_{S}=\pm1$
when moving between triplet LEs and singlet CTs, and $\Delta M_{S}=0$
when moving between singlet CTs and triplet GSs. Actually, we already
knew this from the previous section: The first ISC occurs between
SOMOs and LUMOs (or HOMOs) of opposite monomers and only the latter
has a node at the dimer linkage {[}Fig. \ref{fig:soc-benzyl}c{]}.
Thus, this process occurs predominantly via next-nearest-neighbour
SOCs, which we found before to exhibit $\Delta M_{S}=\pm1$ selectivity
{[}Fig. \ref{fig:soc-benzyl}b{]}. Similarly, the second ISC connects
between two SOMOs of opposite monomers and is mediated mostly by nearest-neighbour
SOCs with $\Delta M_{S}=0$ selectivity {[}Fig. \ref{fig:soc-benzyl}b{]}.
Yet, the key improvement achieved by this section is the perfection
of ISC spin selectivities to $100\%$ by making the opposite spin
channel symmetry-forbidden. From a microscopic perspective, we have
destructively interfered any deleterious ISC channels using other
higher-nearest-neighbour SOCs, treated more generally by the group-theoretical
approach. Importantly, this electronic structure produces the ODMR
mechanism shown in Fig. \ref{fig:main}c, where populations in the
triplet $M_{S}=\pm1$ LEs are moved into the triplet $M_{S}=0$ GS
via two sequential ISCs of opposite spin selectivities. In other words,
such molecules exhibit the ODMR mechanism of NV centres.

As an aside, while it is tempting to consider only symmetry arguments
and ignore the discussions made in the previous section, note that
symmetry-allowed transitions need not occur with high amplitudes.
Therefore, one still needs to check if the allowed ISCs are facilitated
by either nearest- or next-nearest-neighbour SOCs, which have the
largest amplitudes. For example, the \emph{meta}-connected diradicals
explored in our previous work \citep{Poh2024} have negligible SOCs
between singlet CTs and triplet GSs {[}Fig. \ref{fig:main}b{]} because
they are dominated by third-nearest-neighbour effects with a maximum
SOC amplitude of $0.02\mathrm{\,cm^{-1}}$ {[}Fig. \ref{fig:soc-benzyl}b{]}.
In other words, our plan for enhanced ODMR contrasts requires \emph{para}-connectivity
between the two benzylic radicals, consistent with our group-theoretical
analysis (\emph{meta}-connectivity does not yield $D_{2}$ point group
symmetry). We note in passing that vibronic perturbations to the SOC
operator, which are not considered by this study, can modify the spin
selectivity as well \citep{Li2024-2,Baryshnikov2017}. This will be
explored in our upcoming studies.

\subsection*{Our proposal}

\begin{figure*}
\emph{\includegraphics[width=1\textwidth]{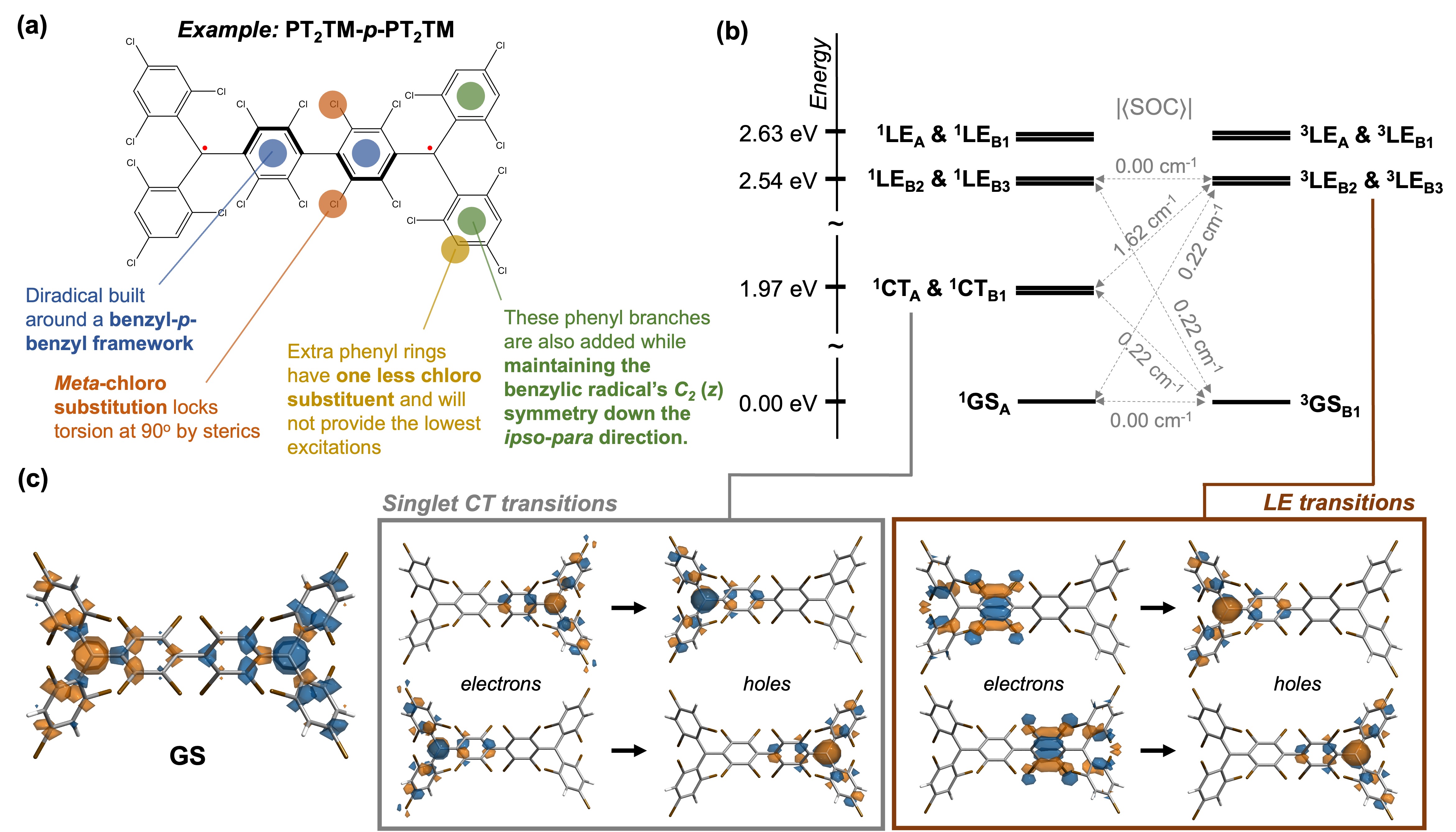}\caption{\label{fig:PT2TM}(a) Features of our diradical design, demonstrated
on the PT\protect\textsubscript{2}TM-\emph{p}-PT\protect\textsubscript{2}TM
diradical. (b) Energy level diagram of PT\protect\textsubscript{2}TM-\emph{p}-PT\protect\textsubscript{2}TM,
computed using TDDFT. Also included are the root-mean-squared SOC
matrix elements, averaged over all channels. These calculations were
performed using MCSCF/CI methods on the central tetra-chlorinated
benzyl-\emph{p}-benzyl fragment. (c) Ground state spin density (isovalue
= 0.004) and natural transition orbitals (isovalue = 0.040) for the
first few excited states of PT\protect\textsubscript{2}TM-\emph{p}-PT\protect\textsubscript{2}TM,
obtained using BS-DFT/TDDFT.}
}
\end{figure*}

We now introduce a molecular design that realises the ODMR mechanism
described in Fig \ref{fig:main}c:$
\global\long\def\theenumi{\arabic{enumi}}%
$
\begin{enumerate}
\item \textbf{The $\pi$-diradical should be a dimer centralised around
a benzyl-}\textbf{\emph{p}}\textbf{-benzyl-like framework.} This ensures
that the diradical's electronic structure resembles Fig. \ref{fig:soc-benzyl}d.
\item \textbf{Sufficient steric hindrances should be installed around the
dimer linkage to maintain a $90^{\text{o}}$ torsion.} Otherwise,
\emph{para} connectivity facilitates $\pi$-bonding and encourages
a singlet ground state \citep{Abe2013,Casado2017,Stuyver2019,Murto2022,Yang2015}.
For instance, chloro substituents can be introduced at the \emph{meta}
positions, which is common practice when synthesising luminescent
benzyl-type radicals \citep{Murto2022}.
\item \textbf{Any additional $\pi$-conjugation to the benzylic radical
should continue to localise the lowest-lying excitations around the
benzylic group.} For instance, phenyl substituents added to the benzylic
radical (typically used to improve radical stability and luminescence)
should have less-extensive internal $\pi$-conjugation so that their
lowest transitions are energetically separated from the benzylic fragment's.
\item \textbf{These extra substituents should also maintain the monoradical's
$C_{2}\br z$ symmetry, where the $z$-axis lies along the }\textbf{\emph{ipso}}\textbf{-}\textbf{\emph{para}}\textbf{
direction.} The reason is that the $C_{2}\br z$ axis is the \emph{only}
symmetry element discriminating between the monomeric $\pi$ MOs {[}Fig.
\ref{fig:soc-benzyl}c{]}, thus their parities under $C_{2}\br z$
dictate the spin selectivity of their intermonomer ISCs. (For the
same reason, diradicals of $D_{2}$ point group symmetry are not the
only molecules satisfying our design criteria. Non-dimeric diradicals
can also be constructed with $C_{2}$ symmetry and the above near-degenerate
pairs of $\curbr{A,B_{1}}$ and $\curbr{B_{2},B_{3}}$ excitations
will become split in energy, thereby offering a ladder of states through
which the singlet molecules can decay. Here, we will focus solely
on dimeric diradicals of $D_{2}$ symmetry for simplicity.) As an
example, when substituting the two hydrogens on the benzylic centre,
both hydrogens should be replaced simultaneously and with identical
moieties.
\end{enumerate}
These features are summarised in Fig. \ref{fig:PT2TM}a using our
prototype PT\textsubscript{2}TM-\emph{p}-PT\textsubscript{2}TM diradical
as an example. Note that its alternacy symmetry is not crucial to
our design, which should be distinguished from our earlier work \citep{Poh2024}.

\subsection*{Ab initio calculations of the prototype PT\protect\textsubscript{2}TM-\emph{p}-PT\protect\textsubscript{2}TM
diradical}

To support the above theory, calculations were performed on the PT\textsubscript{2}TM-\emph{p}-PT\textsubscript{2}TM
diradical as a representative example of experimentally-feasible diradicals.
Starting with its ground state properties, the triplet equilibrium
geometry was found using unrestricted DFT to display a significant
torsion of $86.4^{\text{o}}$ around the dimer linkage. At this geometry,
the ground singlet state is only $1.08\text{ meV}$ lower in energy
(equivalent to thermal energy at $12.5\text{ K}$) than the ground
triplet state, computed using broken-symmetry (BS) DFT. Thus, the
\emph{m}-chloro substituents sufficiently minimise $\pi$-bonding
between the two radicals, that is, we have obtained a true diradical.

Excited state properties were then estimated using unrestricted time-dependent
DFT (TDDFT) within the Tamm--Dancoff approximation (TDA). The electronic
structure fits our requirements for the ODMR mechanism {[}Fig. \ref{fig:main}c{]}
with the exception of another low-lying LE pair of irreps $A$ and
$B_{1}$ at around $0.1\text{ eV}$ away from the lowest LEs {[}Fig.
\ref{fig:PT2TM}b{]}. As eluded in Fig. \ref{fig:main}c, these excitations
are spectroscopically distinguishable from the lowest LEs by wavelength
and may be avoided by optically pumping with a single-frequency laser.
Otherwise, there is a possibility of avoiding these excitations by
aligning the excitation polarisation along the lowest LEs' transition
dipole moments, i.e. along the $xy$-plane. \emph{Even if these undesired
LEs were photoexcited,} any deleterious $\Delta M_{S}=0$ ISCs to
the lowest singlet CTs are expected to be slower than internal conversions
(ICs) to the lowest LEs due to Kasha's rule \citep{Englman1970} and
will have a minimal impact on the proposed ODMR pathway. Further verification
of this energy gap was achieved by applying multi-configurational
self-consistent field (MCSCF) and configuration interaction (CI) methods
on the monoradical. Here, the dimer linkage was capped with a hydrogen
substituent. This approach places the energy separation between the
two lowest LEs at $0.21\text{ eV}$, consistent with the TDDFT/TDA
predictions {[}Fig. \ref{fig:PT2TM}b{]}.

As speculated, the lowest LE transitions are largely localised on
the central (tetra-chlorinated) benzyl-\emph{p}-benzyl fragment {[}Fig.
\ref{fig:PT2TM}c{]}. Hence, for computational efficiency, the SOC
matrix elements were computed for this central fragment instead, capping
any open ends with hydrogen. Indeed, the MCSCF/CI results suggest
appreciable SOC matrix elements along the proposed ODMR pathway {[}Fig.
\ref{fig:PT2TM}b{]}. While, strictly speaking, these calculations
present only an upper bound to the true SOC matrix elements of the
full PT\textsubscript{2}TM-\emph{p}-PT\textsubscript{2}TM diradical,
their values should not decrease by much upon weak $\pi$-conjugation
to the additional trichlorophenyl substituents. Most importantly,
these calculations lend credence to the theoretical proposal presented
by this work.

\subsection*{Symmetry avoids Zeeman-induced singlet-triplet mixings during EPR
spectroscopy}

Interestingly, in connecting two identical monoradicals of (at least)
$C_{2}$ point group symmetry to form a diradical of (at least) $D_{2}$
point group symmetry, the resulting structure can avoid undesirable
singlet-triplet mixings found in many weakly-coupled diradicals under
an applied magnetic field. This would alleviate most magnetic-field-induced
decoherence events in the ground state \citep{Atherton1993,Chizhik2014}
(notice that the singlet and triplet GSs differ by either the coherence
or a single spin flip \citep{Poh2024}). It also facilitates Rabi
nutation experiments in EPR spectroscopy \citep{Kopp2024}, amongst
other benefits.

To introduce the problem, we consider the spin Hamiltonian $\hat{H}_{\text{spin}}$
of two electron spins $\hat{{\bf S}}_{j}$ ($j=1,2$) under an applied
magnetic field ${\bf B}$: 
\begin{align}
\hat{H}_{\text{spin}} & =\mu_{\text{B}}{\bf B}\cdot{\bf g}_{1}\cdot\hat{{\bf S}}_{1}+\mu_{\text{B}}{\bf B}\cdot{\bf g}_{2}\cdot\hat{{\bf S}}_{2}+2J\hat{{\bf S}}_{1}\cdot\hat{{\bf S}}_{2}\nonumber \\
 & \quad+\frac{\mu_{0}g_{\text{e}}^{2}\mu_{\text{B}}^{2}}{4\pi}\sqbr{\frac{\hat{{\bf S}}_{1}\cdot\hat{{\bf S}}_{2}}{\abs{{\bf r}}^{3}}-\frac{3\br{\hat{{\bf S}}_{1}\cdot{\bf r}}\br{\hat{{\bf S}}_{2}\cdot{\bf r}}}{\abs{{\bf r}}^{5}}}.\label{eq:H_spin-local}
\end{align}
The first two terms represent each spin's interaction with the magnetic
field (i.e. the spin Zeeman terms), characterised by different $g$-tensors
${\bf g}_{j}$ ($j=1,2$). The third term symbolises the exchange
interaction between the two spins with coupling $2J$ and is responsible
for singlet-triplet splittings. The last term denotes the spin-spin
dipolar interaction that gives rise to zero-field splittings in organic
diradicals; here, the spatial degrees of freedom (characterised by
${\bf r}$) have been integrated over some orbital subspace (for instance,
the ground state). Finally, the symbols $\mu_{\text{B}}$, $\mu_{0}$,
and $g_{\text{e}}$ represent, respectively, the Bohr magneton, the
vacuum magnetic permeability, and the electron $g$-factor. Implicit
to the above expression is the definition of a four-dimensional Hilbert
space spanned by tensor products of the two electron spin states.
By applying the angular momentum sum rules, one may also express the
Hilbert space in the basis of one singlet and three triplet spin states
\citep{Szabo1989,Atherton1993}. This amounts to re-expressing Eq.
(\ref{eq:H_spin-local}) in terms of the total spin operator $\hat{{\bf S}}\equiv\hat{{\bf S}}_{1}+\hat{{\bf S}}_{2}$,
which yields 
\begin{align}
\hat{H}_{\text{spin}} & =\mu_{\text{B}}{\bf B}\cdot{\bf g}_{1}\cdot\hat{{\bf S}}+\mu_{\text{B}}{\bf B}\cdot\br{{\bf g}_{2}-{\bf g}_{1}}\cdot\hat{{\bf S}}_{2}\nonumber \\
 & \quad+J\br{\hat{{\bf S}}^{2}-\hat{{\bf S}}_{1}^{2}-\hat{{\bf S}}_{2}^{2}}+\hat{{\bf S}}\cdot{\bf D}\cdot\hat{{\bf S}}.\label{eq:H_spin-delocalised}
\end{align}
The procedure for transforming the last term of Eq. (\ref{eq:H_spin-local})
into the last term of Eq. (\ref{eq:H_spin-delocalised}) is presented
in most EPR texts \citep{Atherton1993,Chizhik2014} with ${\bf D}$
being referred to as the zero-field splitting tensor. We now see the
problem: If ${\bf g}_{1}\neq{\bf g}_{2}$, true in most organic systems,
then the singlet-triplet basis block-diagonalises every term in Eq.
(\ref{eq:H_spin-delocalised}) (with the triplets forming their own
block) except for the second one. In other words, the second term
induces mixings between the singlet state and the triplet manifold.
This, as mentioned, leads to deleterious effects such as spin decoherence
\citep{Atherton1993,Chizhik2014} and additional Rabi frequencies
\citep{Kopp2024}. One solution would be to engineer an exchange coupling
$J$ that is large enough for the Zeeman-induced singlet-triplet couplings
of $\sim\mu_{\text{B}}B\abs{g_{2}-g_{1}}$ to be a perturbation relative
to the singlet-triplet gap. However, as explained in our earlier work
\citep{Poh2024}, further enhancement of the exchange coupling in
$\pi$-diradicals is likely to result in a singlet GS, i.e. $J$ is
typically antiferromagnetic \citep{Abe2013,Casado2017,Stuyver2019,Murto2022}.

The alternative approach would be to somehow make elements that differ
between ${\bf g}_{1}$ and ${\bf g}_{2}$ disappear under experimental
conditions. This can be achieved in our diradicals by aligning the
magnetic field ${\bf B}$ parallelly to the dimer linkage. (We acknowledge
that aligning the molecules, say, in a liquid crystal, may be an experimental
challenge, although this has been previously demonstrated \citep{Andrienko2018,Eckvahl2023}.)
To see that, we first note that elements of the $g$-tensor have the
following approximate expression, obtained from second-order perturbation
theory in the SOC: \citep{Atherton1993,Chizhik2014}
\begin{align}
g_{ab} & =g_{\text{e}}\delta_{ab}+2\lambda\sum_{\Psi'}\frac{\bra{\Psi}\hat{L}_{a}\ket{\Psi'}\bra{\Psi'}\hat{L}_{b}\ket{\Psi}}{E_{\Psi}-E_{\Psi'}}.\label{eq:g_ab}
\end{align}
Here, $\lambda$ is an effective SOC constant and $\curbr{\ket{\Psi'}}$,
$\curbr{E_{\Psi'}}$ respectively denote the orbital eigenfunctions
and eigenenergies of the molecule's non-relativistic electronic Hamiltonian,
with $\Psi$ being the orbital eigenfunction under consideration (say,
the ground state wavefunction). The indices $a$ and $b$ run over
all three-dimensional coordinates (such as $x$, $y$, and $z$) and
$\hat{L}_{a}$ is the $a$-th component of the orbital angular momentum
operator $\hat{{\bf L}}$. Because each monoradical of our proposed
system is symmetric about $C_{2}\br z$ rotation (using the coordinate
frame defined by Fig. \ref{fig:soc-benzyl}e, i.e. $z$-axis along
dimer linkage), the $g$-tensor must remain invariant under $C_{2}\br z$
rotation. Such a rotation takes $\curbr{\hat{L}_{x},\hat{L}_{y},\hat{L}_{z}}\rightarrow\curbr{-\hat{L}_{x},-\hat{L}_{y},\hat{L}_{z}}$,
hence, using Eq. (\ref{eq:g_ab}), we obtain 
\begin{align}
{\bf g} & \equiv\br{\begin{array}{ccc}
g_{xx} & g_{xy} & g_{xz}\\
g_{yx} & g_{yy} & g_{yz}\\
g_{zx} & g_{zy} & g_{zz}
\end{array}}=\br{\begin{array}{ccc}
g_{xx} & g_{xy} & -g_{xz}\\
g_{yx} & g_{yy} & -g_{yz}\\
-g_{zx} & -g_{zy} & g_{zz}
\end{array}}.
\end{align}
By contradiction, it must be that 
\begin{align}
{\bf g} & =\br{\begin{array}{ccc}
g_{xx} & g_{xy} & 0\\
g_{yx} & g_{yy} & 0\\
0 & 0 & g_{zz}
\end{array}}.\label{eq:g_1}
\end{align}
Next, the two monoradicals in the dimer are related by a $C_{2}\br y$
rotation. Taking Eq. (\ref{eq:g_1}) to be the $g$-tensor of radical
1, we find the $g$-tensors of both radicals to be
\begin{align}
{\bf g}_{1} & =\br{\begin{array}{ccc}
g_{xx} & g_{xy} & 0\\
g_{yx} & g_{yy} & 0\\
0 & 0 & g_{zz}
\end{array}},\\
{\bf g}_{2} & =\br{\begin{array}{ccc}
g_{xx} & -g_{xy} & 0\\
-g_{yx} & g_{yy} & 0\\
0 & 0 & g_{zz}
\end{array}}.
\end{align}
Therefore, because both ${\bf g}_{1}$ and ${\bf g}_{2}$ have the
same $g_{zz}$ components, the second term of Eq. (\ref{eq:H_spin-delocalised})
vanishes when the magnetic field ${\bf B}$ falls along the $z$-axis
(the dimer linkage), i.e. 
\begin{align}
\hat{H}_{\text{spin}} & =\mu_{\text{B}}g_{zz}\abs{{\bf B}}\hat{S}_{z}\nonumber \\
 & \quad+J\br{\hat{{\bf S}}^{2}-\hat{{\bf S}}_{1}^{2}-\hat{{\bf S}}_{2}^{2}}+\hat{{\bf S}}\cdot{\bf D}\cdot\hat{{\bf S}}.\label{eq:H_spin-B_z}
\end{align}
The singlet-triplet states may now block-diagonalise $\hat{H}_{\text{spin}}$
with no singlet-triplet mixings present. Furthermore, due to the approximate
$D_{2d}$ point group symmetry of our system, the zero-field splitting
caused by ${\bf D}$ also occurs along the $z$-axis. This may be
further rationalised by the cylindrical spin density distribution
of the diradical \citep{Chizhik2014}. Therefore, the singlet-triplet
basis fully-diagonalises $\hat{H}_{\text{spin}}$ in Eq. (\ref{eq:H_spin-B_z})
when the magnetic field is applied along the dimer linkage.

\subsection*{Predicted ODMR contrasts}

\begin{figure*}
\emph{\includegraphics[width=0.8\textwidth]{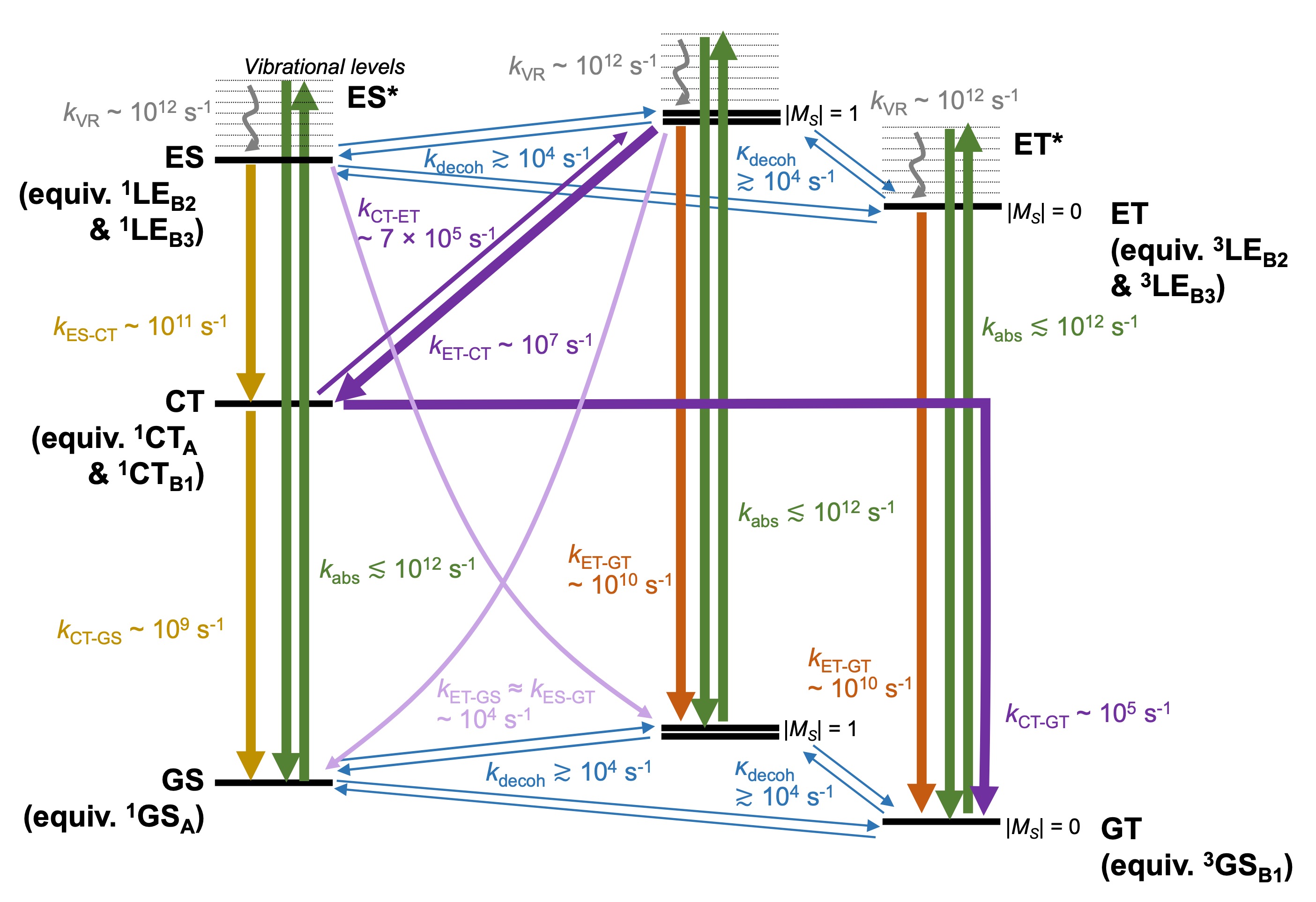}\caption{\label{fig:odmr-a}Schematic diagram depicting the excited-state transitions
being considered by our microkinetic model.}
}
\end{figure*}

As a final theoretical piece to this puzzle, we construct a microkinetic
model of the excited state kinetics and predict the ODMR contrast
at steady state. For simplicity, we shall take the incoherent limit
and consider only changes to the eigenstate populations. Also, populations
of degenerate eigenstates are pooled into a single variable. Therefore,
the relevant eigenstates are $
\global\long\def\theenumi{\alph{enumi}}%
$
\begin{enumerate}
\item the singlet GS (labelled GS, degeneracy = 1),
\item the triplet $M_{S}=0$ GS (labelled GT0 for ground triplet, degeneracy
= 1),
\item the triplet $M_{S}=\pm1$ GS (labelled GT1, degeneracy = 2),
\item the singlet CTs (labelled CT, degeneracy = 2),
\item the singlet LEs (labelled ES for excited singlet, degeneracy = 2),
\item the triplet $M_{S}=0$ LEs (labelled ET0 for excited triplet, degeneracy
= 2), and
\item the triplet $M_{S}=\pm1$ LEs (labelled ET1, degeneracy = 4).
\end{enumerate}
Here, we continue to assume a spin quantisation axis parallel to the
dimer linkage. The kinetic processes connecting these eigenstates
and their associated rate constants are presented in Fig. \ref{fig:odmr-a};
there, the symbols adopt their usual meanings. As far as possible,
model parameters were selected within experimental reason. Listed
below are our justifications for these parameters and the reader may
safely skip it at the first pass. $
\global\long\def\theenumi{\arabic{enumi}}%
$
\begin{enumerate}
\item Also modelled are the vibrationally-excited LE states, labelled with
an asterisk ({*}). These states represent the wavefunction following
an initial Franck-Condon photoexcitation and will, within picoseconds,
undergo vibrational relaxation to the ground vibrational level of
the respective LE states, labelled without an asterisk ( ).
\item The experimentally-observed IC rates from LE to GS triplets are between
$10^{8}$ and $10^{10}\mathrm{\,s^{-1}}$ \citep{Abdurahman2023,Chang2024,Chowdhury2024,Kopp2024}.
In our model, we shall pick the upper bound as a conservative estimate.
Then, because experiments predict the IC rate from CT to GS singlets
to be an order of magnitude slower \citep{Chowdhury2024,Kopp2024},
we place its value at $10^{9}\mathrm{\,s^{-1}}$. Finally, IC from
LE to CT singlets must be faster in view of the smaller energy gap
\citep{Englman1970} and we estimate this rate to be $10^{11}\mathrm{\,s^{-1}}$.
\item The ISC rates are chosen to represent typical organic systems, i.e.
around $10^{7}\mathrm{\,s^{-1}}$ between excited states (i.e. triplet
LEs to singlet CTs) \citep{Kohler2009} and $10^{5}\mathrm{\,s^{-1}}$
for relaxation to the ground (i.e. singlet CTs to triplet GSs) \citep{Kohler2009,Mena2024,Singh2024}.
Because the energy gaps between LEs and GSs are larger than between
CTs and GSs, we expect their ISC rates to be scaled down by an order
of magnitude to around $10^{4}\mathrm{\,s^{-1}}$ (this is also expected
by the El-Sayed rules; see Refs. \citep{Poh2024,Kopp2024}). Lastly,
because the singlet CTs and triplet LEs are relatively close in energy
($\gtrsim0.05\text{ eV}$) \citep{Poh2024,Chowdhury2024,Kopp2024},
reverse ISCs are probable with a delayed timescale. By detailed balance,
this rate will be around $7\times10^{5}\mathrm{\,s^{-1}}$ if we make
a safe assumption of a $0.02\text{ eV}$ energy gap at $85\text{ K}$
-- these parameters also produce similar ODMR contrasts as experiments
\citep{Chowdhury2024,Kopp2024} when the CT-to-GS ISC channel is shut
(more to follow).
\item The weakly-coupled pair of doublet spins in the GS can mix and decohere
which, in the singlet-triplet basis, translates into population transfers
among the singlet and triplet GSs. Therefore, we shall assume a uniform
rate constant $k_{\text{decoh}}$ for mixing between the four near-degenerate
magnetic levels (one from the singlet and three from the triplet).
This is consistent with Kopp et al. \citep{Kopp2024}, where the singlet
to triplet GS ISCs occur at around the same timescale as the triplet
spin-lattice and spin-spin relaxations. The same decoherence pathway
is also expected among the singlet and triplet LEs. Special attention
is given to mixing among the triplet spin sublevels, denoted by a
rate constant $\kappa_{\text{decoh}}$. In our simulations, we will
always set $\kappa_{\text{decoh}}=k_{\text{decoh}}$ unless a microwave
drive is applied (necessary for finding the ODMR contrast). This introduces
additional triplet spin mixing that, assuming a saturating microwave
field, is mathematically equivalent to the limit of $\kappa_{\text{decoh}}\rightarrow\infty$,
i.e. full mixing between the triplet sublevels \citep{Bayliss2020}.
\end{enumerate}
Solving the kinetic model for the steady-state populations yields
the ODMR contrast and optically-induced spin polarisation via the
following expressions \citep{Dreau2011,Tetienne2012,Li2024-2}: 
\begin{align}
 & \text{ODMR contrast}\nonumber \\
 & \quad=\frac{\sum_{j=0,1}n_{\text{ET}j}^{\text{ss},\infty}-\sum_{j=0,1}n_{\text{ET}j}^{\text{ss}}}{\sum_{j=0,1}n_{\text{ET}j}^{\text{ss}}}\times100\%
\end{align}
and 
\begin{align}
\text{Spin polarisation} & =\frac{n_{\text{GT0}}^{\text{ss}}-n_{\text{GT1}}^{\text{ss}}/2}{n_{\text{GT0}}^{\text{ss}}+n_{\text{GT1}}^{\text{ss}}/2}\times100\%,
\end{align}
where the factor of $1/2$ arises from the double degeneracy of the
triplet $M_{S}=\pm1$ GSs. Here, $n_{\Psi}^{\text{ss}}$ denotes the
steady-state population of state $\Psi$ at $\kappa_{\text{decoh}}=k_{\text{decoh}}$,
while $n_{\Psi}^{\text{ss},\infty}$ denotes the same quantity at
$\kappa_{\text{decoh}}\rightarrow\infty$, i.e. under a microwave
resonance. (The rate equations are available in Supplementary Information
1.)

\begin{figure*}
\emph{\includegraphics[width=0.8\textwidth]{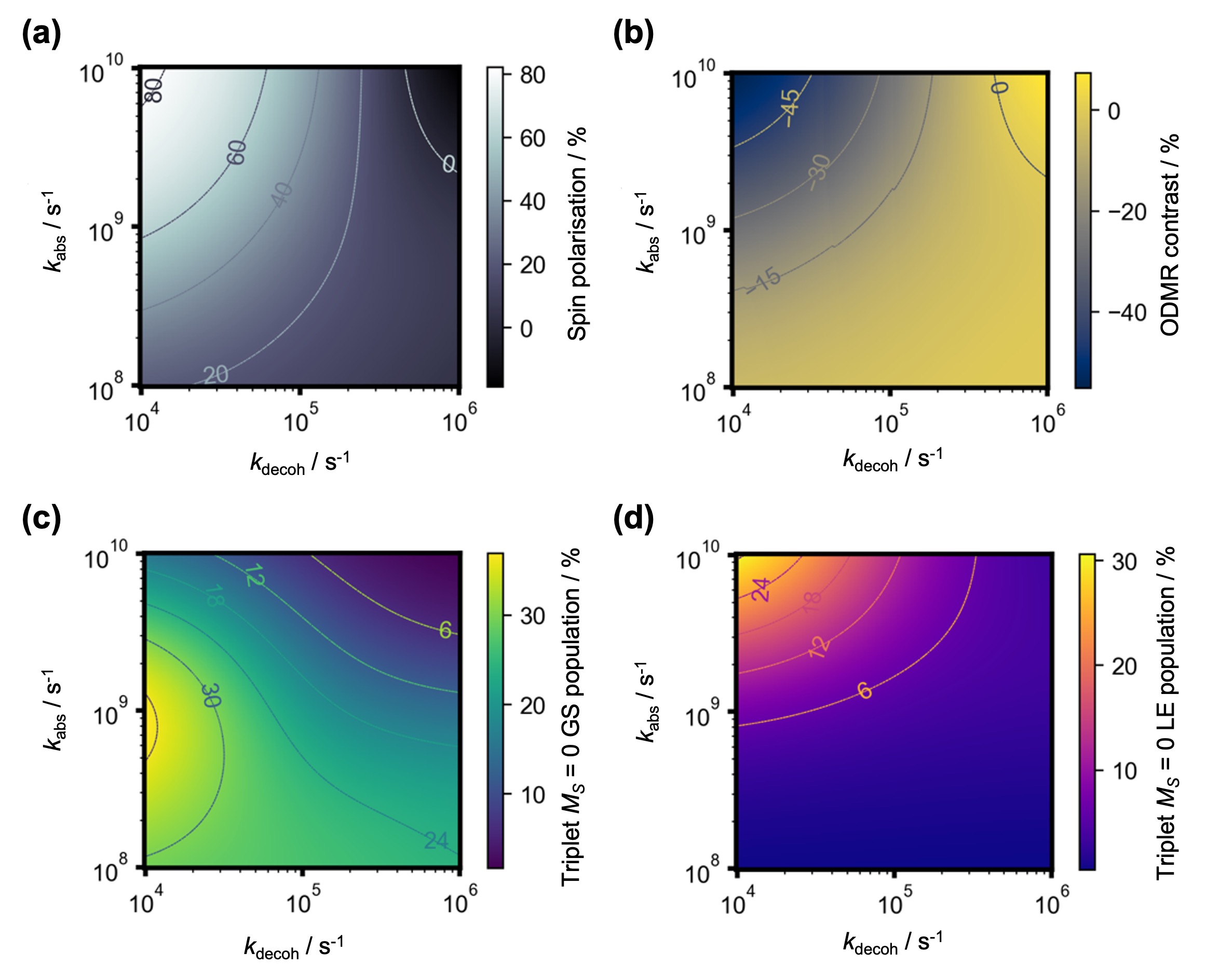}\caption{\label{fig:odmr-b}Steady-state solutions to the (a) optically-induced
spin polarisation, (b) ODMR contrast, (c) triplet $M_{S}=0$ GS population,
and (d) triplet $M_{S}=0$ LE population (which simulates the emission
intensity). Plots are made across varying optical pump rates $k_{\text{abs}}$
and spin decoherence rates $k_{\text{decoh}}$ within experimental
ranges. Other parameters are defined in Fig. \ref{fig:odmr-a}, chosen
to be as realistic as possible.}
}
\end{figure*}

The results are plotted in Fig. \ref{fig:odmr-b} across experimental
ranges of $k_{\text{abs}}$ and $k_{\text{decoh}}$. The former represents
the optical pump rate and has an experimental upper bound of $10^{12}\mathrm{\,s^{-1}}$
for our system (see Supplementary Information 2 for estimates of $k_{\text{abs}}$,
which is based on a laser used in teaching labs \citep{Zhang2018}).
Meanwhile, the latter denotes the spin relaxation rate, the slowest
of which is around $10^{4}\mathrm{\,s^{-1}}$ at temperatures of around
$100\text{ K}$ with faster rates expected at higher temperatures
\citep{Gorgon2023,Schafter2023,Chowdhury2024,Kopp2024}. Strikingly,
at $k_{\text{abs}}=10^{9}\mathrm{\,s^{-1}}$ and $k_{\text{decoh}}=10^{4}\mathrm{\,s^{-1}}$,
a high ODMR contrast of $-27.29\%$ is observed with an optically-induced
spin polarisation of $62.76\%$. The triplet $M_{S}=0$ LE population,
which is proportional to the photoluminescence intensity, is also
appreciable at $7.34\%$. Importantly, the proportion of molecules
in the triplet GSs is above $25\%$, suggesting that the observed
polarisation pathway goes beyond the shelving mechanism described
by previous studies \citep{Poh2024,Chowdhury2024,Kopp2024}. Finally,
the above ODMR contrast was obtained at a microwave drive rate of
$\kappa_{\text{decoh}}\sim10^{6}\mathrm{\,s^{-1}}$, which is experimentally
reasonable \citep{Dreau2011,Tetienne2012,Li2024-2}. For comparison
with our earlier theoretical study \citep{Poh2024}, the predicted
ODMR contrast in the absence of CT-to-GS ISC is only $-0.24\%$, in
agreement with experimental setups that employ the shelving mechanism
\citep{Chowdhury2024,Kopp2024} (this final result was obtained with
$k_{\text{CT-GT}}=0$).

\section*{Conclusion}

By covalently tethering two benzylic radicals at their \emph{para}
positions, the resulting diradical of $D_{2}$ point group symmetry
has a triplet GS of irrep $B_{1}$, lowest singlet CTs of irreps $A$
and $B_{1}$, and lowest triplet LEs of irreps $B_{2}$ and $B_{3}$.
Thus, by group-theoretical considerations, spin polarisation can be
achieved via a $\Delta M_{S}=\pm1$ ISC from LEs to CTs, followed
by a $\Delta M_{S}=0$ ISC from CTs to GSs, just like in an NV centre.
Moreover, the associated SOC matrix elements are appreciable due to
next-nearest-neighbour effects for the former and nearest-neighbour
effects for the latter. We expect this success to be generalisable
to any diradical based upon two \emph{para}-connected benzylic radicals.
This is illustrated in the PT\textsubscript{2}TM-\emph{p}-PT\textsubscript{2}TM
diradical: Firstly, the benzylic radicals are \emph{meta}-chlorinated
to lock the inter-radical torsion by steric repulsion, ensuring that
the GS comprises both triplets and singlets at near-degeneracy. Next,
the stabilising phenyl rings are specially chosen to not break the
$C_{2}$ rotational symmetry down the benzylic \emph{ipso}-\emph{para}
direction, which is the essential symmetry element of the aforementioned
benzylic para-dimer. Finally, every benzylic radical has one more
chloro substituent than each of its phenyl branches, thereby localising
the lowest-lying excitations on the tetra-chlorinated benzylic fragment.
Indeed, these expectations are reflected in the ab initio results
by producing the desired electronic structure for ODMR.

To our best knowledge, our new design has checked most boxes for a
robust optical-spin interface \citep{Wasielewski2020}. Perhaps a
minor problem not addressed by this work is the poor emissive properties
of our PT\textsubscript{2}TM-\emph{p}-PT\textsubscript{2}TM prototype
due to its alternacy symmetry \citep{Abdurahman2020,Hele2021,Poh2024}.
For that, we expect techniques of alternacy symmetry breaking \citep{Abdurahman2020,Hele2021,Gorgon2023}
and excited-state symmetry breaking \citep{Murto2023,Ghosh2024} employed
by the organic light-emitting diode (OLED) community to be useful.
In fact, the latter was recently demonstrated by Chowdhury et al.
\citep{Chowdhury2024}, who synthesised trityl-based diradicals of
near-unity luminescence quantum yields by mesitylating at the \emph{para}
positions. When applied to PT\textsubscript{2}TM-\emph{p}-PT\textsubscript{2}TM,
a possible diradical would look like Fig. \ref{fig:main}c, which
is a suitable synthetic starting point. Notably, the ODMR contrast
for this novel spin polarisation pathway is expected to be around
$30\%$ with more than $25\%$ of the population being in the triplet
manifold at steady state, an unequivocal improvement from earlier
theoretical \citep{Poh2024} and experimental \citep{Chowdhury2024,Kopp2024}
works. Regarding qubit operations, our design also overcomes problems
of Zeeman-induced spin decoherence and observations of multiple Rabi
frequencies during EPR measurements of weakly-coupled diradicals.
These are all crucial steps towards the realisation of optically-addressable
molecular spin qubits.

\section*{Methods}

The geometry of benzyl-\emph{p}-benzyl was relaxed as a triplet on
UB3LYP-D3BJ/def2-SVP \citep{Grimme2010,Grimme2011} while enforcing
$D_{2d}$ point group symmetry. Due to the lack of steric hindrances,
the optimisation converged to a saddle point on the potential energy
surface with a single imaginary frequency of amplitude $57.8\mathrm{\,cm^{-1}}$.
These calculations were done using the ORCA 5.0 code \citep{Neese2022}.
We then computed the SOC matrix elements between carbon $2p$ AOs
of opposite $\pi$-systems using the single-electron form of the SOC
operator \citep{Marian2001}, taking each AO to be singly-occupied
and using a carbon effective nuclear charge of 3.9 \citep{Fedorov2003}.
These integrals were computed using PySCF 2.5.0 \citep{Sun2020} and
the carbon $2p$ AOs were represented by the STO-6G basis set.

Hereafter, all ab initio calculations were performed with the ORCA
5.0 code \citep{Neese2022}. Geometries of PT\textsubscript{2}TM-\emph{p}-PT\textsubscript{2}TM
and its central tetra-chlorinated fragment (labelled Cl\textsubscript{4}M-\emph{p}-Cl\textsubscript{4}M)
were optimised as triplets on UB3LYP-D3BJ/def2-SVP \citep{Grimme2010,Grimme2011}
with harmonic vibrational frequency analyses done to ensure no imaginary
frequencies. The resulting geometries had at least $D_{2}$ point
group symmetries. For the full PT\textsubscript{2}TM-\emph{p}-PT\textsubscript{2}TM
diradical, its electronic structure was estimated by spin-unrestricted
TDDFT/TDA at the UB3LYP/def2-SVPD level using both triplet and BS
singlet ground states as reference. In cases where the latter approach
yielded BS solutions of $\braket{{\bf S}^{2}}\approx1$, we assumed
negligible discrepancies to the singlet energies from spin contamination
because the results were similar to the triplet excitations (within
0.1 eV), which are the most-probable spin contaminants \citep{Yamaguchi1988}.
Excited state irreps were then inferred from the polarisations of
the respective transition dipoles. In performing a BS-DFT calculation,
we had also obtained the open-shell singlet GS energy via the following
expression from Yamaguchi et al. \citep{Yamaguchi1988}: 
\begin{align}
E_{\text{S}}-E_{\text{T}} & =\frac{\braket{{\bf S}^{2}}_{\text{T}}}{\braket{{\bf S}^{2}}_{\text{T}}-\braket{{\bf S}^{2}}_{\text{BS}}}\br{E_{\text{BS}}-E_{\text{T}}},
\end{align}
where $E_{\text{T}}$, $E_{\text{BS}}$, and $E_{\text{S}}$ are the
energies of the triplet, BS singlet, and open-shell singlet states
while $\braket{{\bf S}^{2}}_{\text{T}}$ and $\braket{{\bf S}^{2}}_{\text{BS}}$
are the respective expectations of the total spin-squared. The smaller
Cl\textsubscript{4}M-\emph{p}-Cl\textsubscript{4}M fragment was
then used to compute the SOC matrix elements at the MCSCF/CI level,
also using the def2-SVPD basis set. For that, we first performed a
complete active space self-consistent field (CASSCF) calculation with
10 active electrons and 10 active orbitals, state-averaging the energy
over the ground triplet and singlet states (i.e. SA2). Thereafter,
the excited states were estimated using the complete active space
configuration interaction (CASCI) method, among which the SOC matrix
elements were computed. In this work, the electron spin states are
expressed in the symmetry-aligned coordinate frame shown in Fig. \ref{fig:soc-benzyl}e.
Finally, with the monoradical of PT\textsubscript{2}TM-\emph{p}-PT\textsubscript{2}TM
(labelled PT\textsubscript{2}TM-H), its geometry was first optimised
as a doublet on UB3LYP-D3BJ/def2-SVP \citep{Grimme2010,Grimme2011}
and the presence of a local minimum of $C_{2}$ point group symmetry
was confirmed by harmonic vibrational frequency analysis. Then, using
the def2-SVPD basis set, orbitals were optimised by CASSCF(11,11)
state-specific to the ground doublet, following which the excited
doublet energies were obtained by CASCI. In this last calculation,
the RIJCOSX approximation (RIJCOSX = resolution of identity approximation
for the Coulomb term and chain-of-spheres approximation for the exchange
term) was applied using the def2/JK auxiliary basis set. All CASCI
energies were corrected by the strongly contracted second-order N-electron
valence state perturbation theory (SC-NEVPT2).

\section*{Acknowledgements}

Y.R.P. and J.Y.-Z. were supported through the U.S. Department of Energy
(DOE) under 2019030-SP DOE CalTech Sub S532207 (DE-SC0022089).

\end{document}